# Development of a 3D-CNN-based Prediction Model for Migration Barriers in Plasma-Wall Interactions

Seiki Saito, Keisuke Takeuchi, Hiroaki Nakamura, Yasuhiro Oda,
Kazuo Hoshino, Yuki Homma, Shohei Yamoto, Yuki Uchida

***Abstract*—Understanding the long-term transport of hydrogen isotopes in plasma-facing materials, such as tungsten, is critical for the steady-state operation of magnetic confinement fusion reactors. However, dynamically updating the transition parameters for kinetic Monte Carlo (kMC) simulations as the atomic structure evolves under continuous plasma irradiation remains a severe computational bottleneck. Conventionally, calculating these migration barriers requires the iterative and computationally expensive Nudged Elastic Band (NEB) method. To overcome this limitation, this article presents a highly efficient surrogate model for predicting migration barriers using a three-dimensional Convolutional Neural Network (3D-CNN), establishing the final component necessary to realize on-the-fly molecular dynamics (MD) and kMC hybrid simulations. The proposed deep learning model takes a two-channel volumetric input, the local three-dimensional potential energy distribution and the voxelized spatial coordinates of the initial and final trapping sites, to directly output the migration barrier as a scalar value. Trained on a comprehensive dataset of tungsten-hydrogen configurations evaluated using the Embedded Atom Method (EAM) potential, the model demonstrated robust predictive accuracy, achieving a Mean Absolute Error (MAE) of 0.124 eV and a high coefficient of determination of 0.890. Furthermore, utilizing GPU acceleration, the inference time is reduced to approximately 2.7 milliseconds per barrier, achieving a speed-up ratio of over 23,000 compared to conventional NEB calculations. This extraordinary acceleration effectively resolves the computational barrier of transition rate evaluations, paving the way for large-scale, dynamic modeling of plasma-wall interactions.**

Manuscript received XXXXXXX; revised XXXXXXX; accepted XXXXXXX. Date of publication XXXXXXX; date of current version XXXXXXX. The review of this paper was arranged by Senior Editor XXXXXXX . *(Corresponding author: Seiki Saito.)*

Seiki Saito and Keisuke Takeuchi are with the Graduate School of Science and Engineering, Yamagata University, Yonezawa, Japan (e-mail: saitos@yz.yamagata-u.ac.jp).

Hiroaki Nakamura is with the Department Research, National Institute for Fusion Science, Toki, Japan and Graduate School of Engineering, Nagoya University, Nagoya, Japan (e-mail: nakamura.hiroaki@nifs.ac.jp).

Yasuhiro Oda is with the Simulation Engineering Division, Toyota Technical Development Corporation, Japan (e-mail: yasuhiro.oda42@mail.toyota-td.jp).

Kazuo Hoshino is with the Department of Applied Physics and Physico-Informatics, Keio University, Yokohama, Japan (e-mail: hoshino@appi.keio.ac.jp).

Yuki Homma is with the National Institutes for Quantum Science and Technology, Rokkasho, Japan (e-mail: homma.yuki@qst.go.jp).

Shohei Yamoto is with the National Institutes for Quantum Science and Technology, Naka, Japan (e-mail: yamoto.shohei@qst.go.jp).

Yuki Uchida is with the National Institutes of Technology, Nagaoka College, Nagaoka, Japan (e-mail: yuchida@nagaoka-ct.ac.jp).

***Index Terms*—3D-CNN, Deep learning, Migration barrier, NEB method, Plasma-wall interaction.**

## I. INTRODUCTION

UNDERSTANDING plasma-wall interactions [1, 2] is crucial for achieving steady-state operation of magnetic confinement fusion devices. A key challenge is simulating the long-term behavior of hydrogen isotopes within plasma-facing materials, such as tungsten, whose atomic structures dynamically change under continuous plasma irradiation. To address this, a hybrid computational approach combining molecular dynamics (MD) and kinetic Monte Carlo (kMC) simulations is highly promising [3, 4]. MD is suited for calculating the short-timescale, non-equilibrium injection process of incident particles, while kMC efficiently handles the long-timescale equilibrium diffusion process within the material. However, a major bottleneck in this hybrid approach is the enormous computational cost required to continuously update the kMC input parameters, specifically, the trapping sites and migration barriers, as the atomic structure of the wall evolves dynamically due to plasma exposure.

Conventionally, migration barriers are calculated using techniques such as the Nudged Elastic Band (NEB) method [5] to find the minimum energy path (MEP). While accurate, the iterative nature of the NEB method makes it computationally prohibitive for on-the-fly execution within a large-scale MD-kMC framework. To overcome this limitation, we have been systematically developing deep learning-based surrogate models to accelerate the evaluation of these essential parameters. Our comprehensive roadmap to realize a dynamic-kMC method, capable of responding to dynamic structural evolution, is illustrated in Fig. 1.

This overall strategy consists of three key prediction models, each addressing a specific bottleneck in parameter calculation. First, as indicated by Model-A in Fig. 1, we developed a pix2pix-based deep learning model to rapidly predict the three-dimensional spatial distribution of binding energy directly from a given dynamic atomic configuration [6]. Second (Model-B in Fig. 1), utilizing this predicted energy landscape, we constructed a U-Net-based model to instantly identify the spatial positions of hydrogen trapping sites [7].

In this paper, we present the third and final piece (Model-C) of this framework: a deep learning model for predicting the migration barriers between the identified trapping sites. We propose a 3D Convolutional Neural Network (3D-CNN) that takes the three-dimensional potential energy distribution, and the start and end positions as inputs to directly predict the migration barrier as a scalar value. By completing this final step, we establish a fully accelerated pipeline, capable of reducing parameter calculation costs from minutes to fractions of a second, thereby paving the way for realistic, large-scale hybrid MD-kMC simulations of plasma-wall interactions.

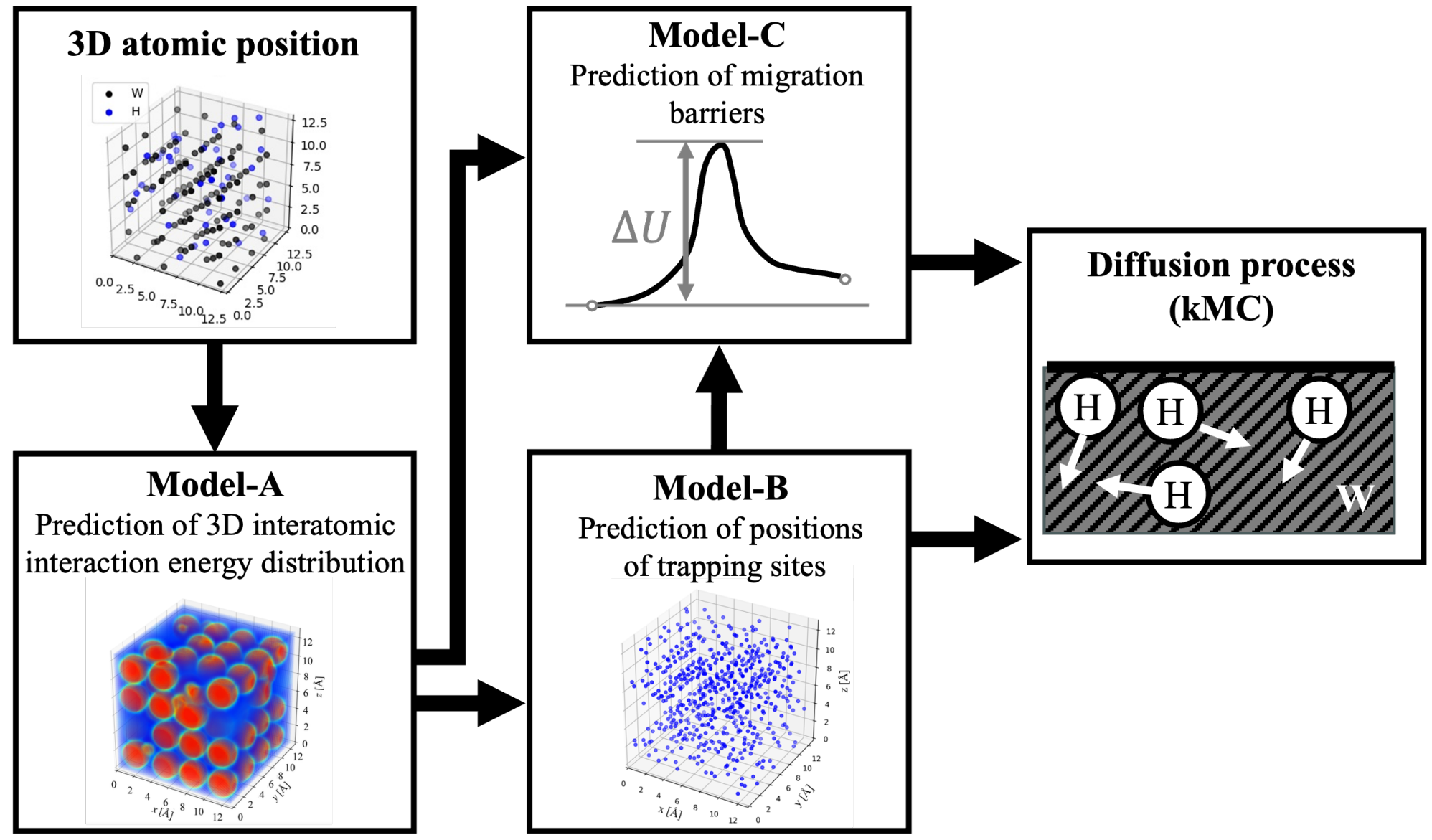

**Fig. 1.** Roadmap for deep learning-based parameter prediction in MD-kMC hybrid simulations. The overall goal is to accelerate the dynamic update of kMC parameters, namely binding energy, trapping sites, and migration barriers, to realize real-time simulation of plasma-facing materials under dynamic structural evolution. This work focuses on the third component (Model-C: Migration Barrier prediction), building upon previous studies (Model-A and B).

## II. Numerical Methodology

### A. NEB Method for Migration Barrier Calculation

In this study, the NEB method is employed to determine the MEP and the corresponding migration barrier when a particle moves from an initial position to a final position within the material, as conceptually illustrated in Fig. 2.

The total potential energy of the system, $U$, which forms the basis for this calculation, is evaluated using the Embedded Atom Method (EAM) potential [8] developed for tungsten-hydrogen interactions. The total potential energy $U$ is defined as:

$$U = \sum_{i} E_i^{\mathrm{emb}} \left[ \sum_{j \neq i} \rho_j(r_{ij}) \right] + \sum_{i,j>i} \phi_{i,j}(r_{ij})$$

where $E_i^{\mathrm{emb}}$ is the embedding energy of atom $i$, $\rho_j$ is the effective electron density contributed by atom $j$, $\phi_{i,j}$ is the pairwise interatomic potential, and $r_{ij}$ is the distance between atoms $i$ and $j$. By utilizing this potential function, the many-body effects of metallic bonding are effectively incorporated, allowing for an accurate evaluation of the energy landscape associated with particle migration.

In the NEB method, multiple virtual atoms are placed at equal intervals between the initial and final states, and they are connected to each other by virtual springs, as shown in the top schematic of Fig. 2. The force $\boldsymbol{F}_i$ acting on the $i$-th virtual atom, representing a potential intermediate state, is decomposed into a perpendicular component $\boldsymbol{F}_i^{\perp}$ and a parallel component $\boldsymbol{F}_i^{\parallel}$ relative to the path:

$$\boldsymbol{F}_i = \boldsymbol{F}_i^{\perp} + \boldsymbol{F}_i^{\parallel}$$

The perpendicular component $\boldsymbol{F}_i^{\perp}$ arises from the true potential gradient of the system (derived from $U$), driving the virtual atom toward the MEP. Conversely, the parallel force $\boldsymbol{F}_i^{\parallel}$ is the artificial spring force that maintains uniform spacing

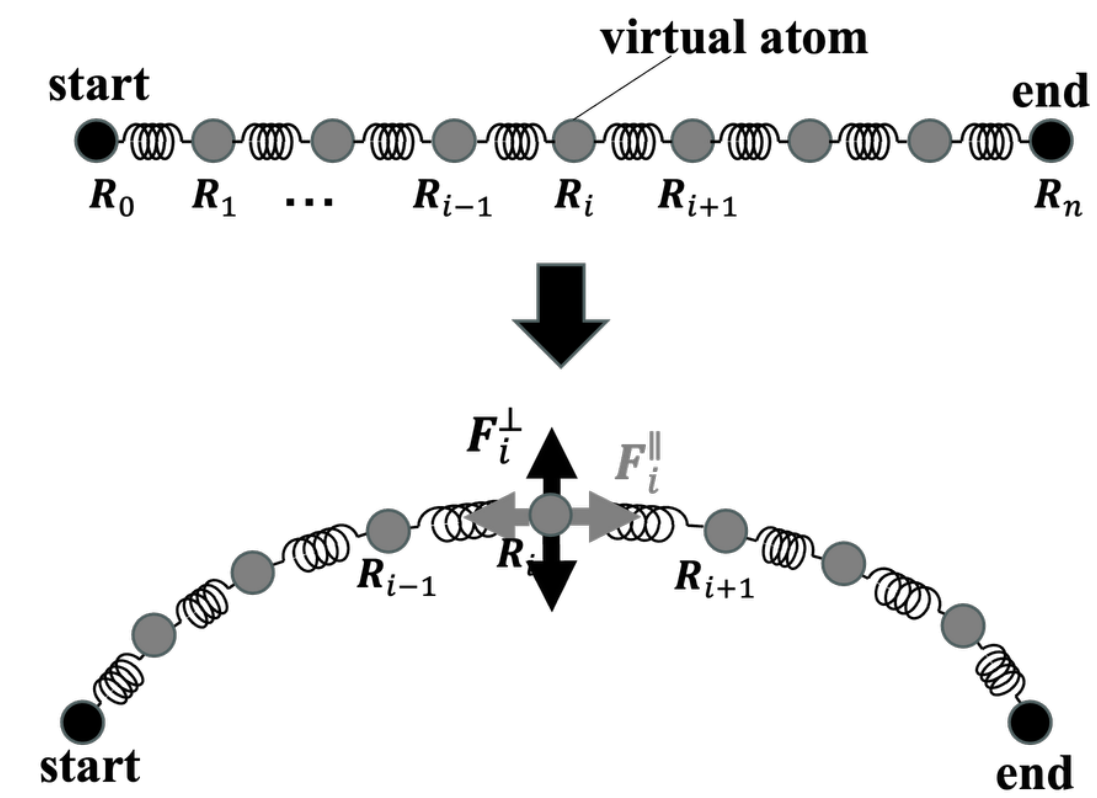

**Fig. 2.** Schematic illustration of the nudged elastic band (NEB) method. A sequence of virtual atoms (images) is connected by artificial springs to form a path between the start and end states. The total force acting on the $i$-th virtual atom is decomposed into the true potential force perpendicular to the path ($\boldsymbol{F}_i^{\perp}$) and the artificial spring force parallel to the path ($\boldsymbol{F}_i^{\parallel}$).

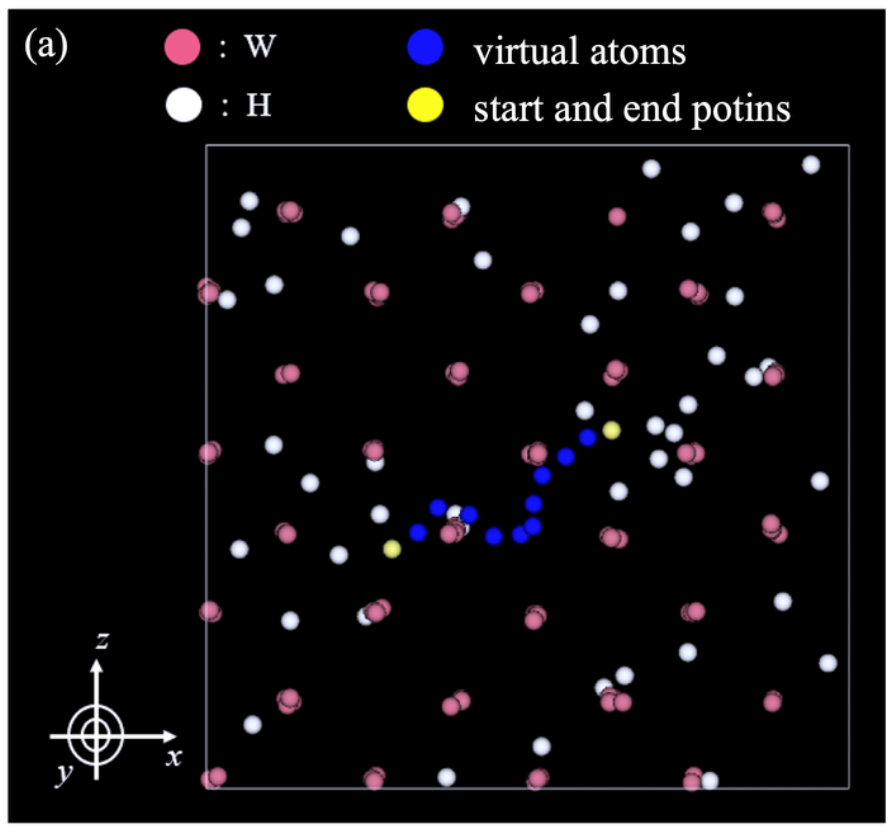

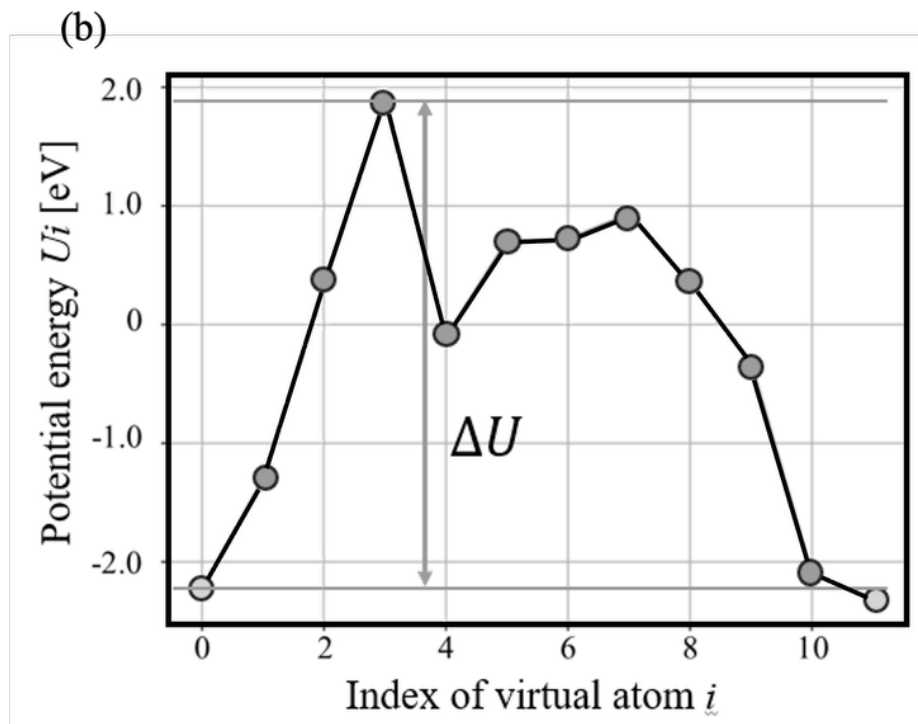

**Fig. 3.** (a) Visualization of the minimum energy path (MEP) calculated by the NEB method in a tungsten-hydrogen system. Blue spheres represent the virtual atoms connecting the start and end points (yellow spheres) within the tungsten lattice (pink spheres) containing pre-existing hydrogen atoms (white spheres). (b) The corresponding potential energy profile along the MEP. The migration barrier $\Delta U$ is defined as the energy difference between the highest saddle point and the initial state.

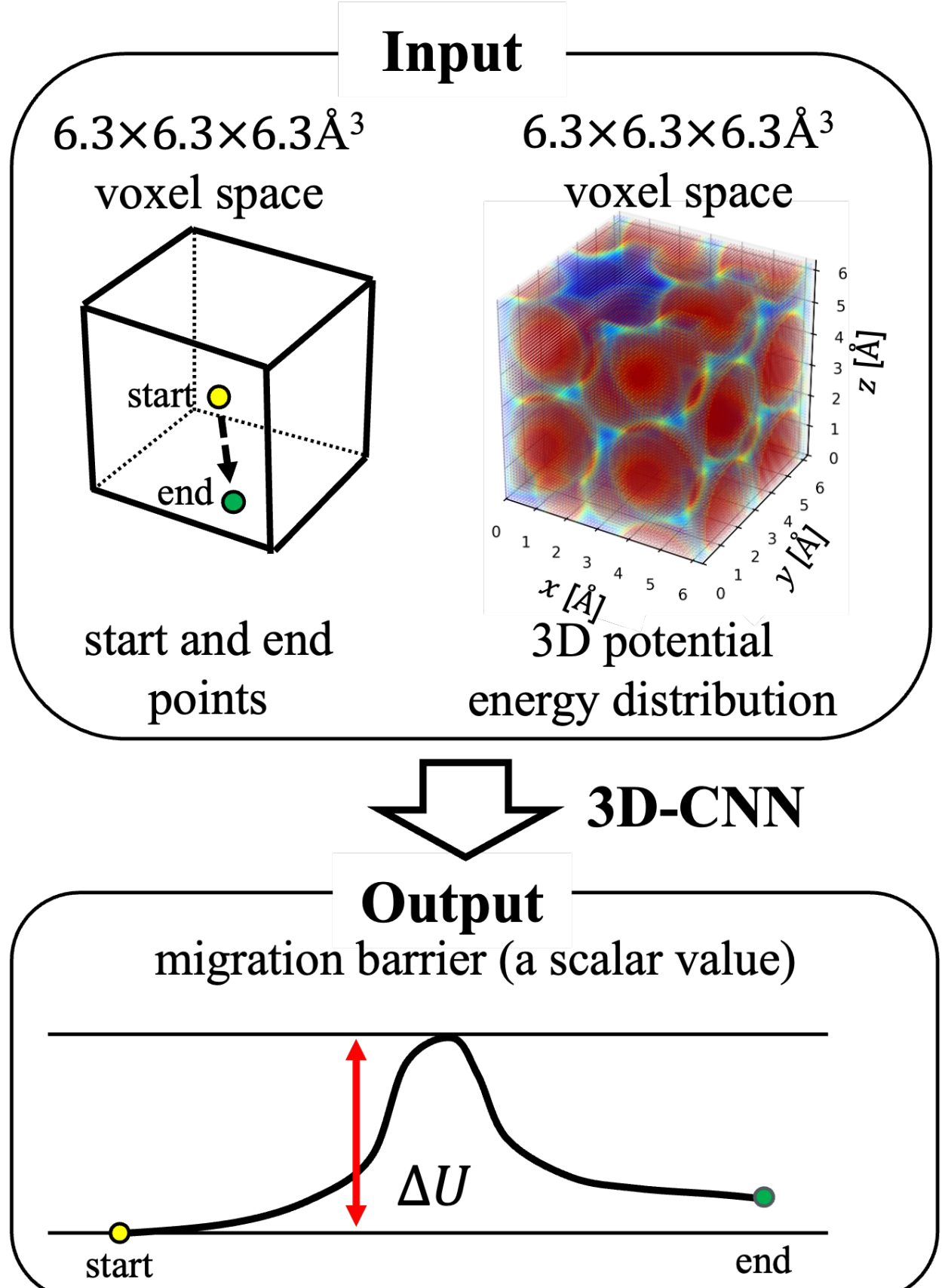

**Fig. 4.** Schematic illustration of the input and output structure of the proposed model. The network takes a two-channel 3D input consisting of the spatial coordinates of the start and end points and the 3D potential energy distribution, and it directly predicts the migration barrier as a single scalar value.

between adjacent virtual atoms along the path. Specifically, these forces are mathematically expressed as:

$$\boldsymbol{F}_i^{\perp} = -\frac{dU}{d\boldsymbol{R}_i} + \left(\frac{dU}{d\boldsymbol{R}_i} \cdot \hat{\boldsymbol{\tau}}_i\right)\hat{\boldsymbol{\tau}}_i\,,$$

$$\boldsymbol{F}_i^{\parallel} = k[(|\boldsymbol{R}_{i+1} - \boldsymbol{R}_i| - |\boldsymbol{R}_i - \boldsymbol{R}_{i-1}|)\hat{\boldsymbol{\tau}}_i]$$

where $\boldsymbol{R}_i$ denotes the position vector of the $i$-th virtual atom, and $k$ is the spring constant. The normalized local tangent vector at the $i$-th virtual atom, $\hat{\boldsymbol{\tau}}_i$, is defined to maintain stability along the path:

$$\hat{\boldsymbol{\tau}}_i = \frac{\boldsymbol{\tau}_i}{|\boldsymbol{\tau}_i|}, \qquad \boldsymbol{\tau}_i = \begin{cases} \boldsymbol{R}_{i+1} - \boldsymbol{R}_i & \text{if } U_{i+1} \geq U_{i-1} \\ \boldsymbol{R}_i - \boldsymbol{R}_{i-1} & \text{if } U_{i+1} < U_{i-1} \end{cases}$$

where $U_i$ represents the potential energy at the $i$-th virtual atom. In this study, the spring constant $k$ is set to 2.2 eV/Å to ensure that these two force components are of comparable magnitude.

Figure 3(a) visualizes an optimized MEP obtained by the NEB method within the tungsten-hydrogen system. The virtual atoms (blue spheres) form a trajectory navigating through the tungsten lattice (pink spheres) and pre-existing hydrogen atoms (white spheres) from the start to the end point (yellow spheres). Once the virtual atoms converge to the MEP, the potential energy $U_i$ of each virtual atom along the path is evaluated. As depicted in the energy profile in Fig. 3(b), the migration barrier $\Delta U$ is determined by the difference between the maximum potential energy along the MEP and the potential energy of the initial state:

$$\Delta U = \max(U_i) - U_0$$

This $\Delta U$ serves as the ground truth for training the proposed 3D-CNN surrogate model.

### *B. 3D-CNN Model Architecture*

To predict the migration barrier directly from the local physical environment, we developed a surrogate model utilizing a 3D-CNN. The schematic illustration of the input and output structure of the proposed model is shown in Fig. 4 and the detailed architecture of the proposed model is illustrated in Fig. 5.

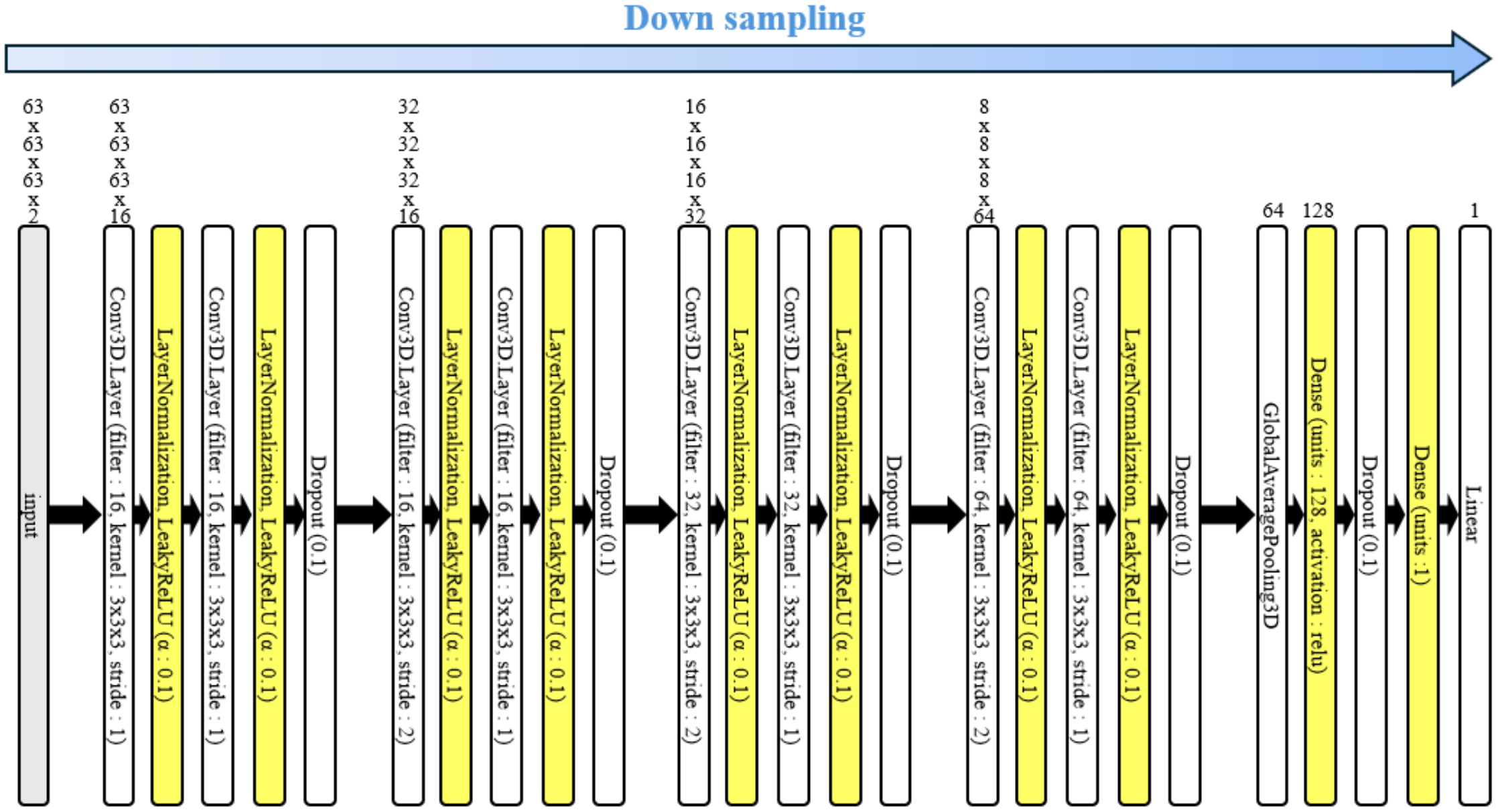

**Fig. 5.** Architecture of the proposed 3D-CNN model for predicting migration barriers. The network processes a two-channel 3D input (positional data and potential energy distribution) through multiple convolutional blocks equipped with Layer Normalization, LeakyReLU, and Dropout, finally outputting a scalar migration barrier value.

The network takes a $63 \times 63 \times 63 \times 2$ tensor as input, which consists of two distinct spatial channels with a spatial resolution of 0.1 Å per voxel. The first channel provides the positional data. The initial and final trapping sites are identified following the method described in Section 3.2, "Identification of trapping sites" of Ref. [7]. In this positional channel, the starting and ending sites are represented by voxels assigned a value of 1, while all other voxels are set to 0. Specifically, the 3D input space is cropped such that the starting site is positioned exactly at the center of the $63 \times 63 \times 63$ voxel grid. Furthermore, the ending site is constrained to be within a distance of 1.58 Å from the start point (i.e., within a spherical shell with a diameter equal to the lattice constant). The second channel contains the three-dimensional potential energy distribution corresponding to this same voxel space. Finally, to focus the model on physically probable events, any data samples with a calculated migration barrier of 5.0 eV or higher were excluded from the dataset.

As shown in Fig. 5, the model employs a down sampling strategy through a series of 3D convolutional blocks to extract hierarchical spatial features from the input data. The network utilizes Conv3D layers with a kernel size of $3 \times 3 \times 3$, progressively increasing the number of filters from 16 to 64. Downsampling is achieved by setting the stride to 2 in specific convolutional layers. To ensure stable training and mitigate overfitting, each convolutional layer is sequentially followed by Layer Normalization, a LeakyReLU activation function with a negative slope coefficient ($\alpha$) of 0.1, and a Dropout layer with a rate of 0.1.

After the spatial dimensions are reduced to $8 \times 8 \times 8$ by the convolutional blocks, a GlobalAveragePooling3D layer is applied to flatten the feature maps. These flattened features are then fed into a fully connected Dense layer consisting of 128 units with a ReLU activation function, followed by another Dropout layer (rate of 0.1). Finally, a Dense output layer with a linear activation function yields a single scalar value. This output represents the predicted migration barrier $\Delta U$ in eV between the specified start and end positions.

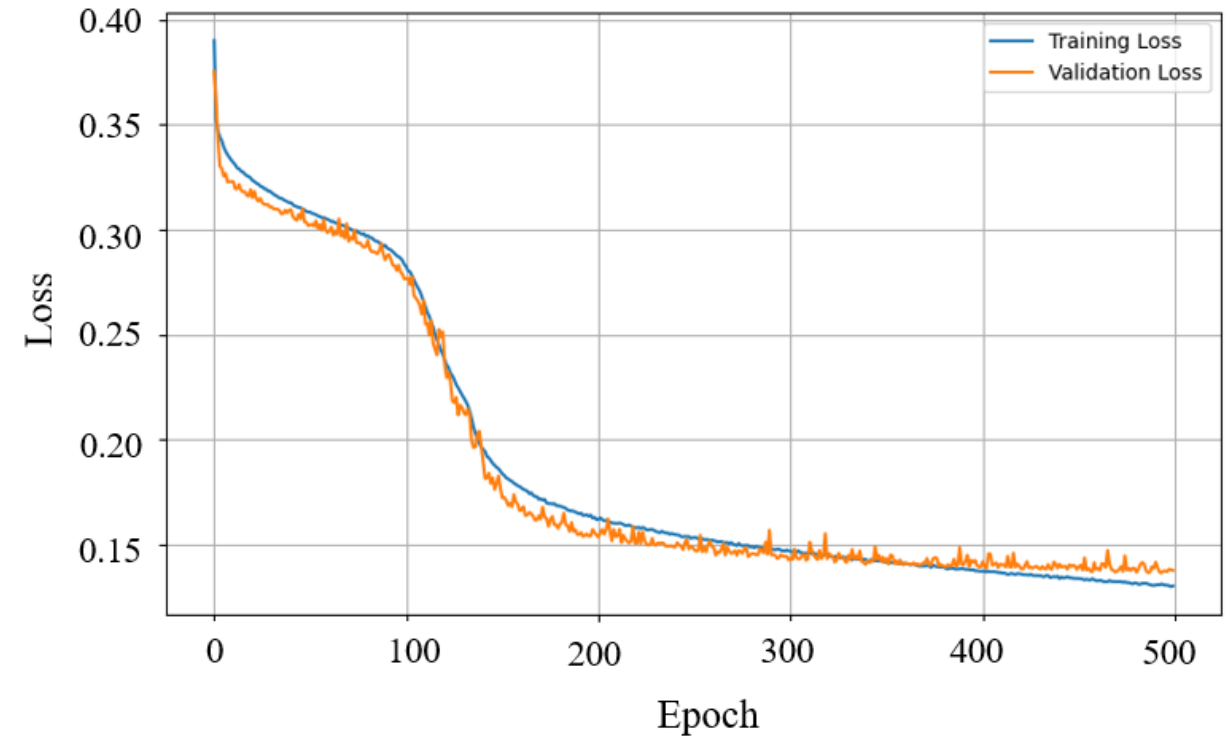

**Fig. 6.** Evolution of the training and validation loss (Mean Absolute Error) over 500 epochs.

The model was implemented in Keras (TensorFlow backend). Training was performed with the Adam optimizer, with an initial learning rate of $1.0 \times 10^{-5}$. All training processes were carried out on an NVIDIA V100 GPU.

## III. Results And Discussion

### *A. Training and Prediction Accuracy*

The 3D-CNN model was trained using a dataset consisting of 82,000 samples. The optimization was performed using the

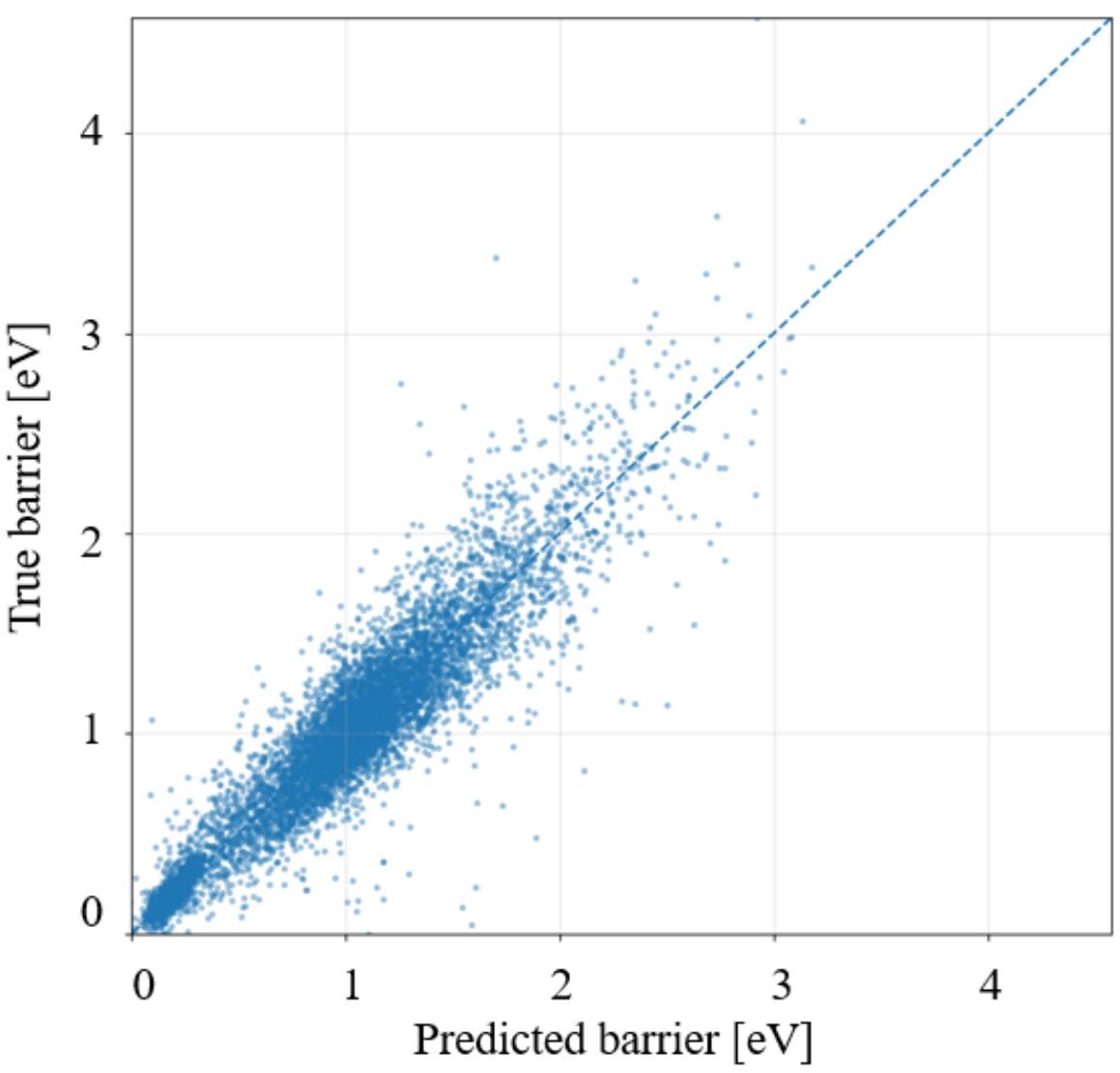

**Fig. 7.** Scatter plot comparing the true migration barriers calculated by the NEB method and the predicted barriers from the 3D-CNN model. The dashed line indicates perfect agreement.

Adam optimizer with a batch size of 32 for a total of 500 epochs. The loss function employed for training was the Mean Absolute Error (MAE).

Figure 6 illustrates the evolution of both the training and validation loss over the 500 epochs. The curves demonstrate a stable and continuous decrease in loss as the training progresses, indicating that the model successfully learns the underlying physical features without significant overfitting.

To quantitatively evaluate the prediction accuracy, the model was tested against unseen data. Figure 7 presents a scatter plot comparing the true migration barriers calculated by the NEB method against the barriers predicted by the 3D-CNN model. The points are densely clustered along the diagonal dashed line, which represents perfect agreement. The model achieved an MAE of 0.124 eV, a Root Mean Square Error (RMSE) of 0.185 eV, and a coefficient of determination ($R^2$) of 0.890. The relatively higher value of the RMSE compared to the MAE suggests that the prediction error tends to increase slightly in regions with larger migration barriers. Nevertheless, the high $R^2$ value confirms that the model is highly capable of predicting migration barriers with an accuracy suitable for practical applications.

*B. Computational Efficiency*

The primary motivation for developing this deep learning surrogate model is to overcome the immense computational bottleneck associated with determining transition parameters for kinetic Monte Carlo (kMC) simulations. Table I summarizes the computational times required to evaluate a single migration barrier using the conventional NEB simulation compared to the proposed 3D-CNN model.

In Table I, $t_{\mathrm{true}}^{\mathrm{1CPU}}$ represents the time required to calculate one migration barrier using the conventional NEB simulation on a single CPU core. The variables $t_{\mathrm{pred}}^{\mathrm{1CPU}}$ and $t_{\mathrm{pred}}^{\mathrm{1CPU+GPU}}$ denote the inference times of the trained 3D-CNN model executed on a single CPU core and with GPU acceleration, respectively.

TABLE I
COMPARISON OF PREDICTION TIMES
FOR SINGLE MIGRATION BARRIER

| | $t_{\mathrm{true}}^{\mathrm{1CPU}}$ | $t_{\mathrm{pred}}^{\mathrm{1CPU}}$ | $t_{\mathrm{pred}}^{\mathrm{1CPU+GPU}}$ |
|---|---|---|---|
| Times (s) | 63.1 | 0.101 | 0.00271 |
| Speed-up ratio | 1 | 623 | 23388 |

As the results indicate, the conventional NEB approach requires approximately 63.1 seconds to evaluate a single barrier ($t_{\mathrm{true}}^{\mathrm{1CPU}}$). In contrast, the 3D-CNN model dramatically reduces this computational expense. Executing the model on a single CPU ($t_{\mathrm{pred}}^{\mathrm{1CPU}}$) takes only 0.101 seconds, achieving a speed-up ratio of 623. Furthermore, when utilizing GPU acceleration ($t_{\mathrm{pred}}^{\mathrm{1CPU+GPU}}$), the prediction time is drastically reduced to just 0.00271 seconds. This corresponds to an extraordinary speed-up ratio of 23,388 compared to the conventional atomistic simulation. This nearly instantaneous prediction capability (approximately 2 ms per barrier) fully satisfies the strict computational time constraints necessary to realize on-the-fly, dynamic MD-kMC hybrid simulations for plasma-wall interactions.

## IV. SUMMARY

In this study, we successfully developed a three-dimensional Convolutional Neural Network (3D-CNN) surrogate model to rapidly and accurately predict the migration barriers of hydrogen within a tungsten lattice. This advancement addresses the most significant computational bottleneck in evaluating transition parameters on-the-fly, which has historically hindered the execution of dynamic kinetic Monte Carlo (kMC) simulations for plasma-wall interactions.

By taking the three-dimensional potential energy distribution and the specific spatial coordinates of the starting and ending trapping sites as inputs, the proposed 3D-CNN model directly outputs the migration barrier as a scalar value. Evaluated against a comprehensive dataset generated using the Nudged Elastic Band (NEB) method, the model demonstrated robust predictive capabilities, achieving a Mean Absolute Error (MAE) of 0.124 eV and a high coefficient of determination ($R^2$) of 0.890.

Most notably, the deep learning approach offers unprecedented computational efficiency. Utilizing GPU acceleration, the inference time per migration barrier is reduced to just 0.00271 seconds, achieving an extraordinary speed-up ratio of 23,388 compared to conventional NEB calculations.

Following our prior successes in utilizing deep learning to predict local potential energy distributions and to identify hydrogen trapping sites, the rapid prediction of migration barriers established in this study provides the final essential component required for dynamic kMC simulations. Future work will focus on integrating these three deep learning models into a unified MD-kMC hybrid simulation framework. This integration will ultimately enable large-scale, long-timescale simulations of plasma-facing materials, providing deeper insights into the dynamic evolution of the tungsten wall under fusion reactor conditions.

## ACKNOWLEDGMENT

The development of the hydrogen trapping site extraction model was supported Grant-in-Aid for Scientific Research No. 22K03572 from the Japan Society for the Promotion of Science, Japan, and by the NIFS Collaborative Research Program (NIFS25KSPT009, NIFS24KIPT013, NIFS25KIST 066, and NIFS22KIGS002). S. Saito is supported by JST (Moonshot R&D Program) Japan Grant Number JPMJMS24A3 for the determination of training data parameter settings based on the wall model specifications. The computations were performed using the Plasma Simulator of NIFS (Toki, Gifu, Japan). The authors would like to acknowledge the use of Gemini (Gemini 3 Flash, Google) for assistance in English language editing and the structural organization of this manuscript.